\documentclass[twocolumn,prd,preprintnumbers,nofootinbib]{revtex4}
\usepackage{graphicx,epsfig}
\usepackage{bm}
\usepackage{latexsym,amssymb,amsmath,float}
\usepackage{color}

\def\deg{\ifmmode{^{\circ}}\else ${^{\circ}}$\fi}
\def\bi{\begin{itemize}}
\def\ei{\end{itemize}}
\def\bfl{\begin{flushleft}}
\def\efl{\end{flushleft}}
\def\ed{\end{document}}

\def\cf#1{\ifmmode{\cal #1}\else${\cal #1}$\fi}

\def\be{\begin{equation}}
\def\ee{\end{equation}}

\def\beas{\begin{eqnarray*}}
\def\eeas{\end{eqnarray*}}
\def\bea{\begin{eqnarray}}
\def\eea{\end{eqnarray}}

\newcommand{\alfnc}{\alpha_{\rm NC}}
\newcommand{\alfcc}{\alpha_{\rm CC}}
\newcommand{\alfbh}{\alpha_{\rm BH}}
\newcommand{\rnc}{r_{\rm NC}}
\newcommand{\rcc}{r_{\rm CC}}

\newcommand{\ie}{i.e.,\ }

\definecolor{rossoCP3}{cmyk}{0,.88,.77,.40}
\newcommand{\nutau}{\nu_\tau}

\begin{document}

\title{New physics with ultra-high-energy neutrinos} 

\author{D. Marfatia$^{1,4}$, D.~W.~McKay$^{2,4}$, and T.~J.~Weiler$^{3,4}$}
\affiliation{$^1$Department of Physics and Astronomy,
University of Hawaii, Honolulu, Hawaii 96822, USA}
\affiliation{$^2$Department of Physics and Astronomy,
University of Kansas,
 Lawrence, Kansas 66045, USA}
\affiliation{$^3$Department of Physics and Astronomy, 
Vanderbilt University, Nashville, Tennessee 37235, USA}
\affiliation{$^4$Kavli Institute for Theoretical Physics, University of California, Santa Barbara, California 93106, USA}

\begin{abstract}
Now that PeV neutrinos have been discovered by IceCube, we optimistically entertain the possibility that neutrinos with energy above 100~PeV exist. We evaluate the dependence of event rates of such neutrinos on the neutrino-nucleon cross section at observatories 
that detect particles, atmospheric fluorescence, or Cherenkov radiation, initiated by neutrino interactions. 
We consider how (i)~a simple scaling of the total standard model neutrino-nucleon cross section, (ii) a new elastic neutral current interaction, and (iii) a new completely inelastic interaction, individually impact event rates.
\end{abstract}


\maketitle

IceCube's announcement of a population of neutrino induced events with shower energies above 1~PeV~\cite{IC1} 
has created excitement in the neutrino astrophysics community.  The long awaited discovery of high energy cosmic neutrinos has arrived. 
Prompted by this discovery, we revisit the problem of extracting neutrino nucleon cross section information from 
currently running and proposed cosmic neutrino  experiments.   
A variety of candidates for sources of the observed neutrinos have been put forward, 
and many ideas for testing models of new physics and old have been advanced, 
but the study of methods to tease out new physics signals from data has not previously gained attention.  
We address this methodology for new physics here, by summarizing the dependence of different detector's acceptance 
of cosmic neutrinos on the cross sections relevant to their propagation and detection.  
We restrict ourselves to ultra-high-energy (UHE) neutrinos, i.e., those with energies above 100~PeV.
Included in ``cross sections'' are any new contributions to neutrino physics.
``Acceptance'' includes all of the calculational factors in the event rate except the flux of incident neutrinos. 

Neutrino detectors naturally segregate into one of three types depending on what aspect of the neutrino-initiated shower is detected: particles,
fluorescence radiation, and radio/visible Cherenkov radiation. 
Particle detectors include Pierre Auger Observatory (PAO)~\cite{auger} and Telescope Array (TA)~\cite{TA}, fluorescence detectors include PAO~\cite{auger}, TA~\cite{TA} and Extreme Universe Space Observatory (EUSO)~\cite{euso},  radio frequency Cherenkov detectors include ANITA~\cite{anita}, ARA~\cite{ara} and  ARIANNA~\cite{arianna}, building on the early searches by the  GLUE~\cite{glue} and RICE~\cite{rice1} experiments,\footnote{The possibility of a phased radio array deployed in glacial ice at Summit Station, Greenland with a PeV-scale threshold is under study~\cite{greenland}.} while the visible Cherenkov detector is IceCube and its expansion to Gen2~\cite{gen2}, which uses deep-ice optical detection. The atmosphere provides the detection medium 
 for PAO, TA and EUSO, while the Antarctic ice provides 
the detection medium for ANITA, ARA , ARIANNA, and IceCube-Gen2.  
Reference~\cite{hmms} was directed specifically to the IceCube configuration.
The geometries of the balloon-borne ANITA and in-ice radio telescopes ARA/ARIANNA, and space-based EUSO make the analyses of the cross section dependence of their event rates quite subtle.  Questions of when a detector ought to be treated as a planar detector or volume detector, or when events are earth-skimming or up-going, and even the effect of surface reflection for in-ice radio telescopes, come into play. 

New physics possibilities naturally segregate into modified total (TOT) cross section, modified neutral current (NC) cross section  (including quasi-elastic when the final state charged lepton 
does not contribute to the shower, as
is the case with produced muons at all energies and produced taus above $\sim 100$~EeV), and enhanced absorption (BH) cross section. A modified total cross section may result from QCD saturation effects or from
new strong interactions like technicolor. An example of a new elastic neutral current-like interaction is provided by enhanced graviton exchange. 
An example of an absorptive enhancement is  
possible micro black-hole production, which is predicted in low scale gravity models.
With this in mind, we label the absorptive enhancement by ``BH''.
By appropriate comparisons between rates of upward and downward going neutrinos in the different experiments (tabulated in Table~\ref{table1}), 
one can isolate the TOT, NC and BH cross section dependences. 
Then, deviations of TOT, NC, or BH cross sections from standard model (SM) expectations would indicate new physics and categorize its potential origin. 

Following Ref.~\cite{hmms}, we parametrize charged current (CC) and NC interactions with the same inelasticity 
(fractional energy transfer to the baryonic target, or $y$ value) as in the SM via
$\alfcc\equiv \sigma_{\rm CC}/\sigma_{\rm TOT}^{\rm SM}$ and
$\alfnc\equiv \sigma_{\rm NC}/\sigma_{\rm TOT}^{\rm SM}$, and parametrize a new
completely inelastic 
cross section (also normalized to $\sigma_{\rm TOT}^{\rm SM}$) by $\alfbh$. 
Then, for the SM, 
($\alfcc, \alfnc, \alfbh)=(\rcc,\rnc,0)\approx (0.71, 0.29, 0)$~\cite{gqrs},
with $\rcc=\sigma_{\rm CC}^{\rm SM}/\sigma_{\rm TOT}^{\rm SM}$ and
 $\rnc=\sigma_{\rm NC}^{\rm SM}/\sigma_{\rm TOT}^{\rm SM}$. 
 A scenario in which the total cross section is scaled by $\alpha$, i.e., $\sigma_{\rm TOT} = \alpha \sigma^{\rm SM}_{\rm TOT}$, 
 is described by ($\alfnc, \alfnc,  \alfbh)=(\alpha \rcc,\alpha \rnc,0$). 
 Similarly, the enhanced NC case with \mbox{$\Delta\sigma_{\rm NC}=\alpha\sigma_{\rm NC}^{\rm SM}$} 
 is described by \mbox{($\rcc,\rnc(1+\alpha),0$)}, and the BH case with $\sigma_{\rm BH}=\alpha\sigma_{\rm TOT}^{\rm SM}$ is described by ($\rcc,\rnc, \alpha$).{\footnote{Note that in Ref.~\cite{hmms}, the NC case is described by \mbox{($\rcc,\rnc+\alpha,0$)} because there $\Delta\sigma_{\rm NC}=\alpha\sigma_{\rm TOT}^{\rm SM}$.}  
In what follows, we distinguish between the attenuation cross section, $\sigma_{\text{att}}$, 
which is relevant for up-going/skimming neutrinos, and the 
showering cross section, $\sigma_{\text{sh}}$. Note that $\sigma_{\rm sh}^{\rm SM}$ is $\sigma_{\rm TOT}^{\rm SM}$ weighted by the energy in the visible shower, 
\ie the total interaction energy minus the non-showering energies of final state neutrinos and
track-producing charged leptons; see Table~\ref{table1}.

The cross section weighted by inelasticity, called the attenuation cross section, for flavor $f$ in the standard model can be written as~\cite{hmms},
\begin{eqnarray}
\sigma^{{\rm SM}f}_{\rm att}
&=&\sigma^{\rm SM}_{\rm CC} 
+\sigma^{\rm SM}_{\rm NC}<y^f_{\rm NC}>\nonumber \\ 
&=&\sigma^{\rm SM}_{\rm CC}+0.2\sigma^{\rm SM}_{\rm NC} \nonumber \\&\simeq& 0.77\sigma^{\rm SM}_{\rm TOT}\,.
\label{att}
\end{eqnarray} 
The attentuation cross sections are the same for the three neutrino flavors (labeled $f=e,\mu,\tau$) because 
\mbox{$<y^f_{\rm NC}> \simeq 0.2$} is the mean inelasticity factor for the NC cross section at energies above 100~PeV~\cite{gqrs}. The final form in Eq.~(\ref{att}) results from the relation $\sigma_{\rm CC}^{\rm SM}\simeq 2.5\sigma_{\rm NC}^{\rm SM}$, independent of energy at UHE for a wide range of cross section estimates~\cite{gqrs}. Note that the attentuation cross section allows for neutrinos that scatter by the NC and continue with 80\% of the original neutrino energy to create a signal in the detector.

For showering in dense media and detection by radio Cherenkov signals at energies above $10^4$~PeV, a first approximation is $\sigma^{\text{\rm SM}}_{\text{sh}}\simeq 0.21\,\sigma^{\text{\rm SM}}_{\text{\rm TOT}}$ for $\nu_e$, $\nu_{\mu}$, and $\nu_{\tau}$, 
with additional contributions from the electromagnetic shower in the $\nu_e$  case, and from $\tau$ decay in matter in the $\nu_{\tau}$ case,
with each new contribution falling with energy.
For the effective showering cross sections, 
factors like the Landau-Pomeranchuk-Migdal (LPM) effect~\cite{lpm} and the  $\tau$ lifetime ($48\,(\frac{E_\tau}{\rm EeV})$~km) 
introduce significant energy dependence into the inelasticity factors~\cite{euso,auger,ara,arianna,rice1}. 

%

First consider the case of  downward neutrino-initiated shower events.  
In the SM, neutrino showers are well-separated 
in the vertical atmosphere
from cosmic-ray showers:
The first interaction of UHE cosmic rays occurs high in the atmosphere ($\sigma_{\rm pN}\sim100$~mb);
on the contrary, UHE neutrinos interact low in the atmosphere, if at all, 
where the atmosphere is exponentially more dense.
For down going neutrinos observed from surface arrays like PAO and TA, or from an airborne observatory like EUSO,
the interaction height ranges from ten meters water equivalent 
for the vertical atmosphere,
to thirty times that for horizontal events~\cite{p-riw}.
The SM neutrino cross section at $10^{20}$~eV is $0.5\times 10^{-31}{\rm cm}^2$, 
and so the optical depth (a measure of the mean number of interactions, or equivalently the interaction probability in the case of an optically thin medium) 
for an incident vertical neutrino is $0.5\times 10^{-4}$,
and  $6\times 10^{-4}$ for an incident horizontal neutrino.
It is unlikely that any new physics cross section would be enormously larger than the SM cross section,
and so we do not anticipate enormously larger optical depths.

%
\vspace*{.1in}
\begin{table*}{}
\begin{tabular}{|c|c|c|c|c|}
\hline
 & & \multicolumn{3}{c|}{}  \\
EXPERIMENT TYPE & SM & \multicolumn{3}{c|}{NEW PHYSICS}  \\ \hline
& & & & \\
   & $\sigma^\text{SM}$ & \ \ $\sigma_{\rm TOT}=\alpha \sigma _{\text{TOT}}^{\rm SM}$\ \  & 
   $\Delta \sigma _{\text{NC}}=\alpha \sigma _{\text{NC}}^{\rm SM}$ & \ \ $\sigma _{\text{BH}} =\alpha\sigma _{\text{TOT}}^{\rm SM}$\ \ \\
 & & & & \\
 \hline
 & & & & \\
 Surface Detector/Fluorescence (in Air): & & & &  \\
 & & & & \\
 PAO/TA/EUSO (down) & $\sigma^\text{{SM}}_{\text{sh}}$ & \ \ \ $\alpha \sigma _{\text{sh}}^{\text{SM}}$ &\ \ \ $\sigma^\text{{SM}}_\text{{sh}}+<y>\Delta\sigma_\text{{NC}}$ \ \ \  
 & \ \ \ $\sigma ^\text{{SM}}_{\text{sh}}+\sigma_{\text {BH}}$   \\
  & & & & \\
 PAO/TA/EUSO{\footnote{For EUSO, the inverse square dependence on the attentuation cross section is an idealization that is somewhat mitigated on detailed modeling~\cite{p-riw}.}}  (up) & \ \ $\frac{ \sigma _{\tau}^{\text{SM}} }{ \left( \sigma _{\text{att}}^{\text{SM}}\right){}^2}$\ \  & \ \ \ $\frac{\alpha \sigma _{\tau}^{\text{SM}}}{\left(\alpha \sigma _{\text{att}}^{\text{SM}}\right){}^2}$ \ \ \
& $\frac{\sigma _{\tau}^{\text{SM}}}{\left(\sigma _{\text{att}}^{\text{SM}}+<y>\Delta \sigma _{\text{NC}}\right){}^2}$ 
& $\frac{\sigma _{\tau }^{\text{SM}}}{\left(\sigma _{\text{att}}^{\text{SM}}+\sigma _{\text{BH}}\right){}^2}$  \\
 & & & & \\
\hline
 & & & & \\
Radio/visible Cherenkov (in Ice): & & & & \\
 & & & & \\
 ANITA/ARA/ARIANNA/IceCube-Gen2 (down)   & $ \sigma _{\rm{sh} }^{\text{SM}}$ &   $\alpha\sigma _{\text{sh}}^{\text{SM}}$  & $\sigma _{\text{sh}}^{\text{SM}}+<y>\Delta\sigma_{\text{NC}}$ & $\sigma^{\text{SM}}_{\text{sh}}+\sigma_{\text{BH}}$ 
  \\
 & & & & \\
 ANITA/ARA/ARIANNA/IceCube-Gen2 (up) &$\frac{ \sigma _{\text{sh}}^{\text{SM}} }{ \sigma _{\text{att}}^{\text{SM}} }$ & $\frac{\alpha \sigma _{\rm{sh} }^{\text{SM}}}{\alpha
\sigma _{\text{att}}^{\text{SM}}}$
&  $\frac{\sigma _{\text{sh}}^{\text{SM}}+<y>\Delta\sigma_{\text{NC}}}{\sigma _{\text{att}}^{\text{SM}}+<y>\Delta \sigma _{\text{NC}}}$ & $\frac{\sigma_{\rm{sh}}^{\text{SM}}+\sigma_{\rm BH}}{\sigma _{\text{att}}^{\text{SM}}+\sigma _{\text{BH}}}$  \\
 & & & & \\
\hline
 \end{tabular}
\caption{
The cross section dependence of the up- and down-going neutrino event rates in the SM and 
in three new physics scenarios for surface/fluorescence and Cherenkov experiments.  
The symbol $\sigma^{\rm SM}_\tau$ stands for the standard $\nutau$-nucleon cross section.
We consider neutrino energies above 100~PeV, so that
the LPM suppression for $\nu_e$ showering in ice above $\sim$~EeV
and $\tau$~escape in air above $\sim100$~EeV, have already occurred.
\label{table1}
}
\end{table*}\vspace{.1in}

A consequence of the same mean inelasticity 
for all flavors is that the NC contribution to the shower signal is flavor-independent. 
The CC flavor cases have different contributions to the shower-calorimetry.  
A $\nu_e$ CC interaction releases 20\% of the energy into a hadron shower and the remaining 80\% into an electromagnetic shower 
as the electron/positron quickly ranges out, so it fully attenuates. Its contribution to showering depends on the medium and the detection method.  
The electromagnetic component contributes fully to the shower detection in air (for PAO, TA and EUSO), 
but the LPM effect in dense media limits its role in generating signal in Cherenkov detectors 
(ANITA, ARA, ARIANNA and IceCube-Gen2) to energies below an~EeV.~\footnote{
For detectors that rely on the radio Cherenkov radiation from showers in ice~\cite{ara, arianna,anita,rice1}, 
the LPM effect causes electromagnetic shower elongation and fluctuation in shower maxima, 
which degrades the coherence of the signal. 
The result is that the dominant mode 
is tau decays into hadrons, for $E_{\nutau}$'s above about 100~PeV.
100~PeV is the threshold for RICE,  and about a factor of ten above the ARA threshold. 
The ANITA and ARIANNA thresholds are above an EeV, so the $\nu_e$ CC contribution is strongly suppressed
and the hadronic tau decays are very dominant.
}

The $\nu_{\mu}$ and $\nu_{\tau}$ collisions, whether CC or NC, transfer only their hadronic recoil portion 
to showers.
However, at energies below $10^{4.5}$ to $10^5$ PeV,  
the $\tau$ produced in a CC $\nu_{\tau}$ interaction decays quickly enough to provide a significant addition to the showers~\cite{sm-t2}.
The detectability of NC events is suppressed because the NC cross section is 2/5 of the CC cross section, and NC events only contain the hadronic shower energy, which at UHE is only 20\% of the incident energy. 
As a first approximation, the highest energy horizontal showers will be all CC $\nu_e$, or totally inelastic, new-physics generated.


For up-going events observable at the Earth's surface, the absorption of the initial neutrino by Earth-matter 
greatly restricts the solid angle of the emerging event. Except for very horizontal events, the Earth is opaque to UHE neutrinos.
In addition, we have seen that the optical depth for a neutrino to interact in our atmosphere is quite small.
Thus, the up going neutrino must interact in the Earth, close enough to the Earth's surface to allow a charged lepton to emerge and shower.
Energy losses for the charged lepton in the Earth, 
and the requirement of a shower, preclude all charged leptons but the tau from emerging and 
showering via its decay~\cite{dhr}.  Thus, up-going neutrinos effecting showers seen above the Earth are restricted to $\nutau$'s.
Remarkably, the rate for up-going, Earth-skimming $\tau$'s from $\nutau$ CC scattering presents an observable signal~\cite{Earth-skimming}.
In fact, the up-going rate scales roughly as $\sigma_{\tau}/\sigma_{\rm att}^2$, due to Earth-absorption effects~\cite{kw}.
The $\tau$'s, of course, emerge almost parallel to the ground. 
This reduced solid angle presents an additional penalty factor for PAO, TA and EUSO~\cite{kw}.
We add that the regeneration effect for $\nutau$'s results in a pile-up of $\nutau$'s at $\sim$~PeV~\cite{Halzen:1998be},
well below the energies of interest to us here.
So we are justified in neglecting $\nutau$ regeneration.


For up-going PAO/TA/EUSO events, a tau lepton produced by an Earth-skimming neutrino collision must emerge into the atmosphere 
still  carrying a substantial fraction of the neutrino energy in order to be detected as an UHE neutrino signal. 
The $\tau$'s relatively small rate of energy loss and short lifetime ($2.9\times 10^{-13}$s in its rest frame) 
allow for a significant chance for detection of its showering decay products by 
 experiments like PAO, TA and EUSO. 
We remark that Earth-curvature effects~\cite{p-riw} become important when the $\tau$ decay length cannot be ignored relative to 
the Earth's radius, i.e., when order $\frac{c\tau_\tau}{R_\oplus} \sim (\frac{E_\tau}{10\,{\rm EeV}})\times 7\%$ accuracy is required.

The upward solid angle available is limited by the ever shorter attenuation length $\lambda_{\rm att}$ as the energy grows.  
The maximum chord length for a neutrino entering the detector volume is 
$\sim \lambda_{\rm att}$, 
so the maximum solid angle is restricted to $2\pi\sin\theta_h$, where
$\sin\theta_h$ = $\lambda_{\rm att}/2R_{\oplus}$ 
is the angle between the entry direction and the horizon. 
Consequently there is a reduction factor of $\lambda_{\rm att}\sim 1/ \sigma_{\rm att}$ in the expected acceptance.
When the $\tau$ lepton must pass below or above the projection area of surface detectors before showering, 
as in the EUSO and PAO/TA experiments, 
this projection carries another $\sin\theta_h$ penalty factor, 
which shows up as the square in the denominators of the PAO/TA/EUSO ``up" rows in Table~\ref{table1}.

On the other hand, the reduced solid angle of the shower in the atmosphere does not affect Cherenkov experiments 
like  ANITA, IceCube-Gen2, ARA and ARIANNA, 
even though the latter two experiments consist of planar, surface detectors.
This is because for these experiments the showers develop in sub-surface ice, thereby enlarging 
the detector volume. 
Thus, for Cherenkov detectors, there is only the single reduction factor in the acceptance, 
$\lambda_{\rm att}\sim 1/ \sigma_{\rm att}$.
Moreover, Cherenkov detection is not limited to $\nutau$ interactions,
but rather to all events that produce showers, regardless of flavor.


Details of the role that SM cross sections play in determining the the acceptance for a given experimental geometry and detection method 
have been elaborated in the literature~\cite{kw,dh,p-riw,hmms,sm-t2}.  
We have drawn on these sources for the comments made above, 
and summarize these comments in the ``SM'' column of Table~\ref{table1}.

Next we turn to the effects of possible new physics.
The case of purely new NC physics, $\Delta\sigma_{\text{NC}}$, adds \mbox{$<y_{\rm NC}>\Delta\sigma_{\text{NC}}$} to the showering cross section.
To estimate the significance of new physics effects in the NC sector, 
we can write the attenuation factor for neutrinos propagating through the Earth as
\begin{equation}
\sigma^{\rm SM}_{\rm {att}}+<y_{\rm NC}>\Delta\sigma_{\rm NC}  \nonumber \\
 = \sigma^{\rm SM}_{\rm CC} 		
	+\,(1+\alpha)\,\sigma^{\rm SM}_{\rm NC}<y_{\rm NC}>\,.  
\end{equation}
Because of the small inelasticity, it 
is seen that an enhancement of $1+\alpha\approx\frac{1}{<y_{\rm NC}>}\sim 5$
is needed to make 
$\sigma^{\Delta {\rm NC}}_{\rm att}$ comparable to $\sigma^{\rm SM}_{\rm TOT}$.


This factor of 5 is relevant for the EUSO experiment, for example.
Downward air showers recorded by EUSO are estimated to receive roughly equal contributions from 
$\nu_e$ CC-initiated showers and $\tau$ decay showers up to 10~EeV, 
but above 100~EeV the $\tau$ showers are a few percent or less because the increased decay length carries the $\tau$ 
outside the observable atmospheric volume before it decays~\cite{p-riw,sm-t2}.  
In Table~\ref{table1}, the cross section dependence for EUSO (down) under $\Delta\sigma_{\rm NC}$ is then $\sigma^{\rm SM}_{\rm CC}$ 
for flavors $e$ and $\tau$ for neutrino energies up to $10^{4.5}$~PeV and for just $e$ above that. 
The contribution of $\Delta\sigma_{\rm NC}$ will be small unless $\alpha\agt 5$. 

Similar considerations lead us to the entries in Table~\ref{table1} for new physics that scales the total SM cross section, 
and for purely inelastic neutrino absorption (BH)}.



In summary, the approximate independence of the up event rate in volume detectors from the total neutrino cross section,
and the fact that
the down event rate is proportional to the flux and the cross section, enables the up/down ratio to isolate the features of the cross section. Since only the deposited energy of interaction is observed, further analysis is needed to link the observed spectrum of events directly to the cross section's dependence on the neutrino energies corresponding to the events.  In the case of surface detectors, the up event rate as a function of the grazing angle can reveal anomalous suppression of up versus down events when new physics is present.  
A known $\nu_{\tau}$ cross section offers an additional handle on the interpretation of the up versus down event rates.  
We believe the overview presented in this paper provides a useful framework to appreciate the general role of the cross sections driving event rates observed in the future.

{\it Acknowledgements.}  This work was supported by the DOE under Grant Nos. DE-SC0010504 and DE-SC0011981, 
and by the Kavli Institute for Theoretical Physics, Santa Barbara (NSF Grant No. PHY11-25915). TJW is also supported by a Simons Foundation Grant, \#306329.

\end{document}